\documentclass[%
 reprint,
twocolumn,
superscriptaddress,
groupedaddress,
 amsmath,amssymb,
 aps,
prb,
]{revtex4}

\usepackage{verbatim}
\usepackage{graphicx}
\usepackage{dcolumn}
\usepackage{bm}
\usepackage{hyperref}

\begin{document}


\title{Hund's coupling stabilized superconductivity in the presence of spin-orbit interactions}

\author{Alfred K. C. Cheung}
\affiliation{%
Department of Physics, Stanford University, Stanford, CA 94305, USA \\
}%


\author{D. F. Agterberg}
\affiliation{
Department of Physics, University of Wisconsin, Milwaukee, WI 53201, USA \\
}%


\date{\today}

\begin{abstract}
The intraorbital repulsive Hubbard interaction cannot lead to attractive superconducting pairing states, except through the Kohn-Luttinger mechanism. This situation may change when we include additional local interactions such as the interorbital repulsion $U^\prime$ and Hund's interactions $J$. Adding these local interactions, we study the nature of the superconducting pairs in systems with tetragonal crystal symmetry including the $d_{xz}$ and $d_{yz}$ orbitals, and in octahedral systems including all three of $d_{xz}$, $d_{yz}$, and $d_{xy}$ orbitals. In the tetragonal case, spin-orbit interactions can stabilize attractive pairing channels containing spin triplet, orbital singlet character. Depending on the form of spin-orbit coupling, pairing channels belonging to degenerate, non-trivial irreducible representations may be stabilized. In the octahedral case, the pairing interactions of superconducting channels are found to depend critically on the number of bands crossing the Fermi energy. 

\end{abstract}

\pacs{Valid PACS appear here}
\maketitle


\section{Introduction}

An on-site repulsive Hubbard interaction cannot, by itself, lead to attractive pairing except through the Kohn-Luttinger mechanism~\cite{raghu1,kohn1}. However, the on-site Hubbard $U$ is not the only local interaction present. When multiple orbitals on each site are considered, then additional atomic interactions are possible, such as the interorbital repulsion $U^\prime$ and Hund's $J$ interaction~\cite{dagotto1}. It is natural to ask whether such additional local atomic interactions can lead to attractive pairing states and thus provide a local mechanism for superconductivity~\cite{spalek1,han1,klejnberg1,hoshino1,werner1,vafek1}. 

A recent work addressing this question through dynamical mean field theory (DMFT) finds that the Hund's interaction can stabilize spin triplet pairing states with electrons residing on different orbitals~\cite{hoshino1}. From a DMFT perspective, the authors invoke the effect of ``spin freezing'' to rationalize their results~\cite{werner1,hoshino1}. Furthermore, the DMFT study finds that the superconducting state is most stable when the Hund's spin coupling is strongly anisotropic. This hints at the importance of spin-orbit coupling which was treated only in a phenomenological manner via an anisotropic spin coupling.

Vafek and Chubukov also examined this problem from a weak coupling perspective~\cite{vafek1}. Motivated by iron-based superconductors, they studied a two dimensional problem with two orbitals per site: $d_{xz}$ and $d_{yz}$. They considered the most general normal state Hamiltonian including local spin-orbit interactions and extracted the local Cooper pairs. Their central finding was that when the renormalized, low energy, $J>U^\prime$, an attractive spin triplet, orbital singlet pairing channel with $A_{1g}$ symmetry emerges. Such a renormalization is consistent with the numerical finding of such a pairing state in DMFT~\cite{hoshino1}. This novel pairing channel only exhibits a pairing instability at the Fermi surface \textit{in the presence of spin-orbit coupling}. Henceforth, we refer to this state as the Vafek-Chubukov (VC) state.

Thus, the theoretical work so far suggests a new mechanism for spin triplet superconductivity based on local \textit{interorbital} atomic interactions, which necessarily require a multi-orbital system, and the presence of on-site spin-orbit coupling. This motivates two directions for further study: (1) What happens when non-local spin-orbit coupling is included? (2) Are these pairing channels unique to the two orbital system in tetragonal crystal symmetry, or can they be realized in octahedral systems including all three $T_{2g}$ orbitals?

In this work, we first revisit the two orbital system and confirm the existence of the VC state which survives in the weak coupling limit when on-site spin-orbit coupling is present. We then add non-local spin-orbit coupling which stabilizes additional pairing channels belonging to non-trivial irreducible representations. We discuss the conditions under which these states are stable. Next, we study the three orbital system with the $d_{xz}$, $d_{yz}$, and $d_{xy}$ orbitals in an octahedral crystal environment. In this limit, the pairing channel analogous to the VC state is found to become \textit{repulsive} when higher energy bands are not allowed to cross the Fermi energy. The energetics of other pairing channels are also critically affected by the number of bands allowed to cross the Fermi energy. 

This article is organized as follows. In Section~\ref{sec:normalH}, we use symmetry methods~\cite{vafek1,fernandes1} to construct the most general normal state Hamiltonian and add local atomic interactions to it to construct the full model. In Section~\ref{sect:tetra}, we study the pairing channels of the two orbital case in tetragonal crystal symmetry and discuss the crucial role played by the form of spin-orbit coupling in determining which channels can be stabilized, as well as the effects of allowing only one band to cross the Fermi energy. In Section~\ref{sect:octa}, we turn to the three orbital case in octahedral crystal symmetry and highlight how constraining bands to not cross the Fermi energy also prevents certain pairing channels from becoming attractive. Section~\ref{sec:summary} contains a summary and further discussions.

\section{Model} \label{sec:normalH}

\subsection{In $D_{4h}$ crystal symmetry}

We first derive the normal state Hamiltonian for the case of two orbitals on each site, $d_{xz}$ and $d_{yz}$. We require the system to have $D_{4h}$ point group symmetry. The most general form of the normal state Hamiltonian $\hat{H}_N$ may be written as:
\begin{equation}
\hat{H}_N=\sum_{\alpha,\beta}c^\dag_{\alpha}\mathcal{H}_{\alpha\beta}c_{\beta},
\label{eq:1}
\end{equation}
in which $\alpha,\beta$ index both orbital and spin quantum numbers. In general, $\mathcal{H}$ may be expressed as a linear combination of direct products of Pauli matrices $\sigma_\mu,\mu=0,...,3$ parametrizing operators in spin space and Pauli matrices $\tau_\mu,\mu=0,...,3$ parametrizing operators in orbital space:
\begin{equation}
\mathcal{H}=\sum_{\mu\nu}f_{\mu\nu}(\vec{k})\sigma_\mu\otimes\tau_\nu,
\label{eq:2}
\end{equation}
in which $f_{\mu\nu}(\vec{k})$ represent general $\vec{k}$ dependent coefficients. In orbital space, $\tau_0$ has $A_{1g}$ symmetry, $\tau_1$ has $B_{2g}$ symmetry, $\tau_2$ has $A_{2g}$ symmetry, and $\tau_3$ has $B_{1g}$ symmetry. In spin space, $\sigma_0$ is $A_{1g}$, $\sigma_1$ and $\sigma_2$ form an $E_g$ set, and $\sigma_3$ is $A_{2g}$. 

In general, all possible combinations of $\mu$ and $\nu$ should be considered. The expression can be substantially simplified by requiring the normal state Hamiltonian to have time reversal and parity symmetry. Note that $\sigma_\mu$ and $\tau_\nu$ are all even under parity such that the direct product $\sigma_\mu\otimes\tau_\nu$ is always even. The coefficient $f_{\mu\nu}(\vec{k})$ must contain only even powers of $\vec{k}$ in order for each term in Eq.~\ref{eq:2} to be overall even under parity. 

Hence, parity symmetry requires $f_{\mu\nu}(\vec{k})$ to be even under time reversal; $\sigma_\mu\otimes\tau_\nu$ must then be even under time reversal for each term in Eq.~\ref{eq:2} to be overall even under time reversal. $\mathcal{H}$ can now be conveniently divided into two components: 
\begin{equation}
\mathcal{H}=\mathcal{H}_0+\mathcal{H}_{s.o.},
\label{eq:3}
\end{equation}
in which $\mathcal{H}_0$ contains terms with $\tau_\nu$ which are even under time reversal ($\nu=0,1,3$) and $\mathcal{H}_{s.o.}$ contains terms with $\tau_\nu$ which are odd under time reversal ($\nu=2$). The reason for the chosen subscripts will become clear momentarily.

Because the coefficients $f_{\mu\nu}$ and the $\tau_\nu$ appearing in $\mathcal{H}_0$ are by construction even under time reversal, the spin operator is necessarily $\sigma_0$. Thus, $\mathcal{H}$ represents the ``bare'', non-spin-orbit coupled part of the normal state Hamiltonian. We may express $\mathcal{H}_0$ as:
\begin{equation}
\mathcal{H}_0=\sigma_0\otimes[f_{00}\tau_0+f_{01}\tau_1+f_{03}\tau_3].
\label{eq:4}
\end{equation}
The remaining task is to determine the form of the coefficients $f_{0\nu}$. This is accomplished by requiring each term to have $A_{1g}$ symmetry overall. This yields:
\begin{align}
\mathcal{H}_0=\sigma_0\otimes [&(A+Bk_\perp^2+Ck_z^2)\tau_0 \nonumber \\
&+D k_x k_y \tau_1 +E(k_x^2-k_y^2)\tau_3].
\label{eq:5}
\end{align}
in which $k_\perp^2\equiv k_x^2+k_y^2$ and $A,B,C,D$ and $E$ represent material dependent parameters independent of $\vec{k}$.

Following the same line of reasoning, $\mathcal{H}_{s.o.}$ can also be determined. The spin matrices involved here must be $\sigma_3$, $\sigma_2$, and $\sigma_3$ for each term to be even under time reversal. Thus, $\mathcal{H}_{s.o.}$ describes spin-orbit coupling interactions and can be expressed as:
\begin{equation}
\mathcal{H}_{s.o.}=\mathcal{H}_{s.o.}^\Gamma+\mathcal{H}_{s.o.}^{\vec{k}},
\label{eq:6}
\end{equation} 
where:
\begin{equation}
\mathcal{H}_{s.o.}^\Gamma=\alpha\sigma_3 \otimes \tau_2,
\label{eq:7}
\end{equation}
is the local spin-orbit interaction containing $\vec{k}$-independent terms, and:
\begin{align}
\mathcal{H}_{s.o.}^{\vec{k}}=\gamma(k_x k_z \sigma_1\otimes\tau_2+k_y k_z \sigma_2\otimes\tau_2),
\label{eq:8}
\end{align}
represents all non-local spin-orbit interactions. $\alpha$ and $\gamma$ are material dependent parameters independent of $\vec{k}$.

\subsection{In $O_h$ crystal symmetry}

The same steps may be repeated to obtain the normal state Hamiltonian for the three orbital case in crystals possessing $O_h$ point group symmetry. The main difference is that because we work with three orbitals, the orbital space is now parametrized by the nine Gell-Mann matrices $\lambda_\mu,\mu=0,...,8$. Specific forms of the Gell-Mann matrices used and our choice of indices to label the orbitals can be found in Appendix~\ref{appendix}. The symmetry classifications of all matrices under the $O_h$ point group are as follows: $\lambda_0$ has $A_{1g}$ symmetry, $\{\lambda_1,\lambda_2,\lambda_3\}$ have $T_{2g}$ symmetry, $\{\lambda_4,\lambda_5,\lambda_6\}$ have $T_{1g}$ symmetry, and $\{\lambda_7,\lambda_8\}$ have $E_{g}$ symmetry. For the spin Pauli matrices, $\sigma_0$ has $A_{1g}$ symmetry and $\{\sigma_1,\sigma_2,\sigma_3\}$ have $T_{1g}$ symmetry. 

The net effect is to restore the symmetry between the $z$ direction and the $x$ and $y$ directions. We simply state the final result here:
\begin{widetext}
\begin{align}
\mathcal{H}_0^{O_h}=\ &\sigma_0\otimes \left[(A+Bk^2)\lambda_0+E(k_x k_y\lambda_1+k_y k_z \lambda_2+k_x k_z \lambda_3)+F((k_x^2-k_y^2)\lambda_7+(2k_z^2-k_x^2-k_y^2)\lambda_8/\sqrt{3})\right], \\
\mathcal{H}_{s.o.}^{\Gamma,O_h}=\ &\alpha(\sigma_3 \otimes \lambda_4 -\sigma_1\otimes \lambda_5 + \sigma_2 \otimes \lambda_6), \\
\mathcal{H}_{s.o.}^{\vec{k},O_h}=\ &\lambda\left[k_xk_z(\sigma_1\otimes\lambda_4-\sigma_3\otimes\lambda_5)+k_yk_z(\sigma_2\otimes\lambda_4+\sigma_3\otimes\lambda_6)+k_xk_y(\sigma_1\otimes\lambda_6-\sigma_2\otimes\lambda_5)\right] \nonumber \\
 &+\gamma\left[-(k_x^2-k_y^2)(\sigma_1\otimes\lambda_5+\sigma_2\otimes\lambda_6)+(1/3)(2k_z^2-k_x^2-k_y^2)(2\sigma_3\otimes\lambda_4+\sigma_1\otimes\lambda_5-\sigma_2\otimes\lambda_6)\right],
\label{eq:9}
\end{align}
\end{widetext}
in which $k^2\equiv k_x^2+k_y^2+k_z^2$. These replace the corresponding quantities in Eq.~\ref{eq:3} and Eq.~\ref{eq:6}.

\subsection{Adding local atomic interactions} \label{sect:channels}

Finally, we add local atomic interactions $\hat{V}$ to the normal state Hamiltonian so that the total Hamiltonian is:
\begin{equation}
\hat{H}=\hat{H}_N+\hat{V},
\label{eq:12}
\end{equation}
in which~\cite{dagotto1}:
\begin{align}
\hat{V}=&\ \frac{U}{2}\sum_{i\gamma,\sigma\neq\sigma^\prime}n_{i\gamma\sigma}n_{i\gamma\sigma^\prime}+\frac{U^\prime}{2} \sum_{i\sigma,\sigma^\prime,\gamma\neq\gamma^\prime}n_{i\gamma\sigma}n_{i\gamma^\prime\sigma^\prime} \nonumber \\
&+\frac{J}{2}\sum_{i\sigma,\sigma^\prime,\gamma\neq\gamma^\prime}c_{i\gamma\sigma}^\dag c_{i\gamma^\prime \sigma^\prime}^\dag c_{i\gamma\sigma^\prime}c_{i\gamma^\prime\sigma} \nonumber \\
&+\frac{J}{2}\sum_{i\sigma\neq\sigma^\prime,\gamma\neq\gamma^\prime}c_{i\gamma\sigma}^\dag c_{i\gamma\sigma^\prime}^\dag c_{i\gamma^\prime\sigma^\prime}c_{i\gamma^\prime\sigma}.
\label{eq:13}
\end{align}
Here, $i$ is the site index, $\gamma,\gamma^\prime$ are orbital indices, and $\sigma,\sigma^\prime$ are spin indices. The first term describes repulsion between antiparallel electrons on the same orbital, parametrized by $U>0$. It is expected to be the dominant energy scale. The second term describes repulsion among electrons on different orbitals, parametrized by $U^\prime>0$. The third and fourth terms respectively represent the Hund exchange interaction and the Hund ``pair hopping'' interaction, both parametrized by $J>0$. The goal is to extract the energies and forms of the local Cooper pairs that arise from these local atomic interactions compatible with the relevant normal state Hamiltonian.

\section{Tetragonal system with two orbitals}\label{sect:tetra}

We first examine the two orbital system with tetragonal point group symmetry considered in Ref.~\onlinecite{vafek1}. Without spin-orbit coupling, six pairing channels emerge. Their pairing interactions, symmetries, and gap structures are summarized in Table~\ref{table:1}. 

\begin{center}
\begin{table}[ht]
\begin{tabular}{|c|c|c|c|}
\hline
 & Energy & Irrep. & Structure \\
\hline\hline
1.\ &$U^\prime -J$ & $A_{1g}$ & $\tau_2\otimes\sigma_3\sigma_2$  \\
2.\ &$U^\prime -J$ & $E_g$    & $\tau_2\otimes\sigma_1\sigma_2$  \\
3.\ &$U^\prime -J$ & $E_g$    & $\tau_2\otimes\sigma_2\sigma_2$  \\
4.\ &$U        +J$ & $A_{1g}$ & $\tau_0\otimes\sigma_0\sigma_2$ \\
5.\ &$U^\prime +J$ & $B_{2g}$ & $\tau_1\otimes\sigma_0\sigma_2$  \\
6.\ &$U        -J$ & $B_{1g}$ & $\tau_3\otimes\sigma_0\sigma_2$  \\
\hline
\end{tabular}
\caption{Local Cooper pair channels due to atomic interactions in the absence of spin-orbit coupling for the two orbital system in tetragonal crystal symmetry.  Channels 1, 2, and 3 are the novel ``spin triplet, orbital singlet'' pairs that can be stabilized when $J>U^\prime$. Channel 1 is the VC state.}
\label{table:1}
\end{table}
\end{center}

We focus first on Channel 1 whose Cooper pair takes the form $\tau_2\otimes\sigma_3(i\sigma_2)$ (where $\tau_\mu$ describes the pairing in orbital space and $\sigma_\nu(i\sigma_2)$ describes the pairing in spin space). Channel 1 represents a pair with $A_{1g}$ symmetry because $\tau_2$ and $\sigma_3$ both have $A_{2g}$ symmetry and $A_{2g}\otimes A_{2g}=A_{1g}$. Channel 1 corresponds to the VC channel. Crucially, its pairing interaction may become attractive when $J>U^\prime$.


What are the conditions that must be satisfied for a given pairing channel to develop a non-zero $T_c$ once spin-orbit coupling is turned on? In the weak coupling limit, a superconducting channel will lead to a pairing instability only if it involves \textit{intraband pairing}. The intuition for this can be seen by the following argument. Begin in the limit when all orbitals have the same kinetic energy $\epsilon(\vec{k})$. In this case, different orbitals have the same Fermi surface, just like spin up and spin down electrons in conventional $s$-wave superconductors in the absence of a Zeeman field. Consequently, an orbitally anti-symmetric Cooper pair will be stable in the weak coupling limit in the same way a spin singlet state is stable. Now consider adding some energy splitting between the different orbitals -- this will look like a Zeeman field in the spin analogy. Once this splitting becomes of the order of the gap, the weak coupling state will no longer survive. In general, different bands will be split off from one another, thus induced intraband pairing is required to avoid the obstacle to orbitally anitisymmetric superconductivity described above. 

A theoretical formalism for determining whether a pairing channel develops intraband pairing as required for a pairing instability in the weak coupling instability was developed in Refs.~\onlinecite{ramires1} and~\onlinecite{ramires2}. The key result was the introduction of the concept of ``superconducting fitness'', quantities which may be calculated to determine whether intraband and/or interband pairing are present. The relevant quantity for determining whether intraband pairing is present is:
\begin{equation}
F_A(\vec{k})(i\sigma_2)=\mathcal{H}(\vec{k})\Delta(\vec{k})+\Delta(\vec{k})\mathcal{H}^*(-\vec{k}),
\label{eq:fitness1}
\end{equation}
where $\Delta(\vec{k})$ is the gap matrix of interest (last column in Table~\ref{table:1}) and $\mathcal{H}(\vec{k})$ is the normal state Hamiltonian matrix. \textit{If $F_A(\vec{k})(i\sigma_2)\neq 0$, then an intraband pairing component exists}. For the VC state with $\Delta(\vec{k})=\tau_2\otimes\sigma_3(i\sigma_2)$, all kinetic energy terms of the normal state Hamiltonian do not contribute to $F_A(\vec{k})(i\sigma_2)$. However, the local spin-orbit interaction $\alpha\tau_2\otimes\sigma_3$ guarantees an non-zero $F_A(\vec{k})(i\sigma_2)$:
\begin{align}
F_A(\vec{k})(i\sigma_2)\sim&\ (\tau_2\otimes\sigma_3)(\tau_2\otimes\sigma_3(i\sigma_2)) \nonumber \\
&+(\tau_2\otimes\sigma_3(i\sigma_2))(\tau_2\otimes\sigma_3)^* \nonumber \\
\sim&\ (\tau_2\otimes\sigma_3)(\tau_2\otimes\sigma_3(i\sigma_2)) \nonumber \\
&-(\tau_2\otimes\sigma_3(i\sigma_2))(\tau_2\otimes\sigma_3) \nonumber \\
\sim&\ \tau_0\otimes i\sigma_2 - \tau_0 \otimes (-i\sigma_2) \nonumber \\
\sim&\ 2\tau_0 \otimes i\sigma_2 \neq 0.
\label{eq:fitness2}
\end{align}
Thus, spin-orbit coupling is essential for the VC state to develop an intraband pairing component and hence possess a pairing instability at a finite temperature. The kinetic energy terms of the normal state Hamiltonian can be shown to lead to interband pairing. Their existence hence suppresses the $T_c$, but they cannot make $T_c$ go to zero as long as spin-orbit coupling is present. The same result was obtained in Ref.~\onlinecite{ramires2}.

\subsection{Gap equations for the VC state} \label{projectionconsequence}

Note that Channel 4 in Table~\ref{table:1} also has $A_{1g}$ symmetry and is therefore allowed to mix with the VC channel. In fact, the intraband pairing component acquired by the VC channel upon turning on spin-orbit coupling arises precisely due to mixing with Channel 4. Channel 4 is an $s$-wave spin singlet channel and has a repulsive pairing interaction $U+J$. The key result obtained by Vafek and Chubukov in Ref.~\onlinecite{vafek1} is that mixing of the VC channel with Channel 4 does \textit{not} suppress this pairing channel. However, implicit in their work was the assumption that both bands cross the Fermi energy and are relevant for pairing. They do not consider pairing restricted to a single band. We show that in the single band limit, the two channels compete and the resulting pairing channel becomes repulsive.

Let $\Delta_{vc}$ and $\Delta_{ss}$ be the order parameters associated with the VC superconducting channel and the on-site spin singlet repulsive channel, respectively. The result above can be seen most readily from the coupled gap equations obeyed by $\Delta_{vc}$ and $\Delta_{ss}$ when only the local spin-orbit coupling is included in the normal state Hamiltonian:
\begin{align}
-\frac{\Delta_{vc}}{g_{vc}}&=\sum_{\vec{k}}[f(\epsilon_+) (\Delta_{vc}+\Delta_{ss})+f(\epsilon_-) (\Delta_{vc}-\Delta_{ss})]\nonumber \\
-\frac{\Delta_{ss}}{g_{vc}}&=\sum_{\vec{k}}[f(\epsilon_+) (\Delta_{vc}+\Delta_{ss})+f(\epsilon_-) (\Delta_{ss}-\Delta_{vc})],
\label{eq:gapeqs1}
\end{align}
where $g_{vc}=U^\prime-J$ and $g_{ss}=U+J$, $f(E)=\tanh(\beta E/2)/E$, and $\epsilon_+$/$\epsilon_-$ are the energies associated with the two bands (they are $\vec{k}$-independent since we are ignoring the $\vec{k}$-dependent terms in the normal state Hamiltonian). In spite of the strongly repulsive $\Delta_{ss}$ channel, Vafek and Chubukov have solved these equations when both bands cross the chemical potential and find that a pairing instability exists for the state with both $\Delta_{vc}$ and $\Delta_{ss}$ non-zero.

Now consider the case when only one band is relevant. Projecting out one of the bands is equivalent to taking $\epsilon_-\rightarrow \infty$. In this limit, $f(\epsilon_-)\rightarrow 0$. Hence, the coupled gap equations reduce to:
\begin{align}
-\frac{\Delta_{vc}}{g_{vc}}&=\sum_{\vec{k}} f(\epsilon_+) (\Delta_{vc}+\Delta_{ss})\nonumber \\
-\frac{\Delta_{ss}}{g_{vc}}&=\sum_{\vec{k}} f(\epsilon_+) (\Delta_{vc}+\Delta_{ss}).
\label{eq:gapeqs2}
\end{align}
It immediately follows that $\Delta_{vs}/g_{vs}=\Delta_{ss}/g_{ss}$. Thus, there is only one distinct gap whose $T_c$ is governed by the manifestly repulsive interaction $g_{ss}+g_{vc}=U+U^\prime$. Hence, we conclude that the result of Vafek and Chubukov arises only in the limit when both bands cross the Fermi energy.

\subsection{The $E_g$ channels} \label{egchann}

In addition to the VC channel, Channels 2 and 3, which comprise a degenerate pair of $E_g$ symmetry, can also be attractive when $J>U^\prime$. These channels have the form $\tau_2\otimes\sigma_1(i\sigma_2)$ and $\tau_2\otimes\sigma_2(i\sigma_2)$ and are hence also orbital singlet, spin triplet states. Note that unlike the VC channel, there are no other competing, repulsive channels of $E_g$ symmetry. Hence, their pairing interaction always remains $U^\prime-J$. Evaluation of the superconducting fitness function $F_A(\vec{k})(i\sigma_2)$ is therefore sufficient to determine their stability. 

For these states, the on-site spin-orbit coupling interaction is unable to induce intraband pairing. However, non-local spin-orbit interactions of the form in Eq.~\ref{eq:8} can induce intraband pairing in these states. Once these interactions are included, we find that the non-trivial, \textit{even parity} $E_g$ channels are generically stable states exhibiting spin triplet, orbital singlet superconductivity. Such terms were neglected in the treatment in Ref.~\onlinecite{vafek1} because they considered a two dimensional limit. Experimental evidence suggests that the $E_g$ pairing channel is realized for URu$_2$Si$_2$~\cite{schemm1,yano1,kasahara1,ikeda1}. Here, multiple orbitals are indeed relevant to the electronic bands near the chemical potential~\cite{ikeda1}, suggesting that the argument presented here may provide a viable explanation for the stability of this state. 

The remaining three channels in Table~\ref{table:1} are all clearly repulsive in their interactions (assuming $U>J$ for Channel 6).

\section{Octahedral system with three orbitals}\label{sect:octa}

We now consider the octahedral case when all three of the $d_{xz}$, $d_{yz}$, and $d_{xy}$ orbitals are included. We work near the $\Gamma$ point and assume that the on-site spin-orbit interaction $\mathcal{H}_{s.o.}^{\Gamma,O_h}$, parametrized by $\alpha$, is the dominant energy scale. Under this assumption, the eigenstates of the normal state Hamiltonian split into a twofold degenerate higher energy manifold (pseudospin $j=1/2$) and a fourfold degenerate lower energy manifold (pseudospin $j=3/2$). 

Initially, we consider the local pairing channels that emerge after restricting to the $j=3/2$ manifold by projecting out the $j=1/2$ manifold. Under these assumptions, six pairing channels emerge from the local atomic interactions. Their energies and symmetries are summarized in Table~\ref{table:2}. We label these six channels with letters A to F to avoid confusion with the channels of the two orbital case of Table~\ref{table:1}.

\begin{center}
\begin{table}[ht]
\begin{tabular}{|c|c|c|}
\hline
 & Energy & Irrep. \\
\hline\hline
A.\ & $2U/3+U^\prime/3+J$ & $A_{1g}$ \\
B.\ & $U^\prime -J/3$     & $T_{2g}$ \\
C.\ & $U^\prime -J/3$     & $T_{2g}$ \\
D.\ & $U^\prime -J/3$     & $T_{2g}$ \\
E.\ & $U/3+2U^\prime/3-J$ & $E_g$    \\
F.\ & $U/3+2U^\prime/3-J$ & $E_g$    \\
\hline
\end{tabular}
\caption{Local Cooper pair channels due to atomic interactions in three orbital octahedral system. Channel A is the state analogous to the VC channel in the two orbital case.}
\label{table:2}
\end{table}
\end{center}

\subsection{What happens to the VC state?}

In the octahedral, three orbital case, the state analogous to the VC state has the form $\vec{\lambda}\cdot\vec{\sigma}(i\sigma_2)$ where we define $\vec{\lambda}=(\lambda_6,\lambda_5,\lambda_4)^T$ and $\vec{\sigma}=(\sigma_1,\sigma_2,\sigma_3)^T$. At first, it may be tempting to associate the the VC analog with one of the $T_{2g}$ channels found (Channels B, C, and D), as these have energies $U^\prime-J/3$ and can therefore be potentially attractive. However, this cannot be the case because the VC state and its three orbital analog has $A_{1g}$ symmetry. The VC analog must instead be associated with Channel A. Channel A however contains a strongly repulsive contribution $2U/3$ in its pairing interaction and cannot be attractive. 

Analogous to the discussion for the tetragonal two orbital case in Section~\ref{projectionconsequence}, this discrepancy can be understood as a result of projecting out the higher energy $j=1/2$ subspace. Without spin-orbit coupling, the VC analog state with pairing interaction $g_{1,A_{1g}}=U^\prime-J$ and the on-site spin singlet state ($\sim\lambda_0 \otimes \sigma_0(i\sigma_2)$) with pairing interaction $g_{0,A_{1g}}=U+2J$ do not mix (all pairing channels in the absence of spin-orbit coupling are summarized in Table~\ref{table:3}). Under the ``single manifold'' limit, the VC analog state forms a linear combination with the on-site spin singlet state and acquires the energy $U$ in its pairing interaction to create a repulsive pairing state. The on-site spin singlet state also belongs to $A_{1g}$ and is allowed to mix with the VC analog state. 

Again, this argument can be made more rigorous by deriving the coupled linearized gap equations for the order parameters of the two channels. Let the combined gap matrix of the system be:
\begin{equation}
\hat{\Delta}=\hat{\Delta}_{0,A_{1g}}+\hat{\Delta}_{1,A_{1g}},
\label{eq:cubic1}
\end{equation}
where $\hat{\Delta}_{0,A_{1g}}=\Delta_{0,A_{1g}}\lambda_0 \otimes \sigma_0(i\sigma_2)$ and $\hat{\Delta}_{1,A_{1g}}=\Delta_{1,A_{1g}} \vec{\lambda}\cdot\vec{\sigma}(i\sigma_2)$. $\Delta_{0,A_{1g}}$ and $\Delta_{1,A_{1g}}$ are the gaps for the on-site spin singlet state and the VC analog state, respectively. The linearized gap equations can be most conveniently obtained through the following expression for the free energy~\cite{agterberg1,ho1}:
\begin{equation}
\mathcal{F}=\frac{1}{2}\sum_{i=0}^1 \frac{1}{g_i}\textrm{Tr}[\hat{\Delta}_i^\dag \hat{\Delta}_i]-\frac{\beta}{2}\sum_{\vec{k},\omega_n} \sum_{\ell=1}^\infty \frac{1}{\ell} \textrm{Tr}[\hat{\Delta}\hat{\tilde{G}} \hat{\Delta}^\dag \hat{G}]^\ell,
\label{eq:cubic2}
\end{equation}
where $\beta=1/k_B T$, $\hat{G}$ and $\hat{\tilde{G}}$ are Green's function matrices associated with the normal state Hamiltonian $\hat{H}_N$, defined by $(i\omega_n - \hat{H}_N)^{-1} \hat{G} = 1$ and $(i\omega_n + \hat{H}_N^T)^{-1} \hat{\tilde{G}} = 1$, and $\omega_n$ are the fermionic Matsubara frequencies $(2n+1)\pi/\beta$. For simplicity, we ignore momentum dependence and take $\hat{H}_N=\epsilon \lambda_0 \sigma_0 + \alpha \vec{\lambda}\cdot\vec{\sigma}$.

\begin{center}
\begin{table}[ht]
\begin{tabular}{|c|c|c|c|}
\hline
Irrep. & Spin structure & Energy & Gap structure \\
\hline\hline
$A_{1g}$ & Singlet  & $U+2J$       & $\lambda_0 \sigma_0(i\sigma_2)$     \\
         & Triplet  & $U^\prime-J$ & $\vec{\lambda}\cdot\vec{\sigma}(i\sigma_2)$ \\
\hline
$T_{2g}$ & Singlet	& $U^\prime+J$ & $\lambda_1 \sigma_0(i\sigma_2)$	\\ 
         &        	&              & $\lambda_2 \sigma_0(i\sigma_2)$	\\ 
         &        	&              & $\lambda_3 \sigma_0(i\sigma_2)$	\\ 
         & Triplet	& $U^\prime-J$ & $(\lambda_6\sigma_2+\lambda_5\sigma_1)(i\sigma_2)$	\\ 
         & 					& 						 & $(\lambda_6\sigma_3+\lambda_4\sigma_1)(i\sigma_2)$	\\ 
         & 					& 						 & $(\lambda_5\sigma_3+\lambda_4\sigma_2)(i\sigma_2)$	\\ 
\hline
$E_{g}$  & Singlet	& $U-J$        & $\lambda_7 \sigma_0(i\sigma_2)$	\\ 
         &        	&              & $\lambda_8 \sigma_0(i\sigma_2)$	\\ 
         & Triplet	& $U^\prime-J$ & $(\lambda_6\sigma_1-\lambda_5\sigma_2)(i\sigma_2)$	\\ 
         & 					& 						 & $(\lambda_6\sigma_1+\lambda_5\sigma_2-2\lambda_4\sigma_3)(i\sigma_2)$ \\ 
\hline
$T_{1g}$ & Triplet	& $U^\prime-J$ & $(\lambda_6\sigma_2-\lambda_5\sigma_1)(i\sigma_2)$	\\ 
         & 					& 						 & $(\lambda_6\sigma_3-\lambda_4\sigma_1)(i\sigma_2)$	\\ 
         & 					& 						 & $(\lambda_5\sigma_3-\lambda_4\sigma_2)(i\sigma_2)$	\\ 
\hline
\end{tabular}
\caption{All fifteen local Cooper pair channels \textit{without spin-orbit coupling}. We define $\vec{\lambda}=(\lambda_6,\lambda_5,\lambda_4)^T$ and $\vec{\sigma}=(\sigma_1,\sigma_2,\sigma_3)^T$. All spin triplet channels have pairing interaction $U^\prime-J$ without spin-orbit coupling. The pairing interactions are derived from the atomic interactions of Eq.~\ref{eq:13}. When spin-orbit coupling is present and the single-manifold approximation is made, channels belonging to the same irreducible representations mix to form the channels in Table~\ref{table:2}.}
\label{table:3}
\end{table}
\end{center}

We take only the $\ell=1$ contribution in the summation over $\ell$ in Eq.~\ref{eq:cubic2}. The two coupled linearized gap equations then follow directly from $\partial \mathcal{F}/\partial \Delta_{i,A_{1g}}=0$ for $\Delta_{0,A_{1g}}$ and $\Delta_{1,A_{1g}}$:
\begin{align}
\frac{6\Delta_{0,A_{1g}}}{g_{0,A_{1g}}}&=\sum_{\vec{k}}[-2(\Delta_{0,A_{1g}}-\Delta_{1,A_{1g}})f(\epsilon_4) \nonumber \\
&-(\Delta_{0,A_{1g}}+2\Delta_{1,A_{1g}})f(\epsilon_2)] \nonumber \\
\frac{12\Delta_{1,A_{1g}}}{g_{1,A_{1g}}}&=\sum_{\vec{k}}[2(\Delta_{0,A_{1g}}-\Delta_{1,A_{1g}})f(\epsilon_4) \nonumber \\
&-2(\Delta_{0,A_{1g}}+2\Delta_{1,A_{1g}})f(\epsilon_2)],
\label{eq:cubic3}
\end{align}
where $\epsilon_4=\epsilon-\alpha$ and $\epsilon_2=\epsilon+2\alpha$ are, respectively, the normal state eigenenergies associated with the $j=3/2$ and $j=1/2$ manifolds. The approximation of projecting out the $j=1/2$ manifold associated with energy $\epsilon_2$ then amounts to taking $\epsilon_2\rightarrow\infty$. In this limit, $f(\epsilon_2)\rightarrow 0$ and the second terms on the right hand sides of Eqs.~\ref{eq:cubic3} can be discarded. Eqs.~\ref{eq:cubic3} become:
\begin{align}
\frac{6\Delta_{0,A_{1g}}}{g_{0,A_{1g}}}&=-\sum_{\vec{k}}2(\Delta_{0,A_{1g}}-\Delta_{1,A_{1g}})f(\epsilon_4) \nonumber \\
\frac{12\Delta_{1,A_{1g}}}{g_{1,A_{1g}}}&=\sum_{\vec{k}}2(\Delta_{0,A_{1g}}-\Delta_{1,A_{1g}})f(\epsilon_4).
\label{eq:cubic4}
\end{align}
Hence, $2\Delta_{1,A_{1g}}/g_{1,A_{1g}}=-\Delta_{0,A_{1g}}/g_{0,A_{1g}}$ implying that only one distinct gap remains after projecting out the upper manifold. This gap obeys the following gap equation:
\begin{equation}
\Delta_{0,A_{1g}} = -\frac{1}{3}\sum_{\vec{k}}(2g_{0,A_{1g}}+g_{1,A_{1g}})\frac{\Delta_{0,A_{1g}}}{2\epsilon_4}\tanh(\beta \epsilon_4/2),
\label{eq:cubic5}
\end{equation}
and is governed by the pairing interaction $(2g_{0,A_{1g}}+g_{1,A_{1g}})/3=2U/3+U^\prime/3+J$, exactly the energy found for the channel of $A_{1g}$ symmetry in Table~\ref{table:2}. 

We conclude that the analog of the VC state in the cubic, three orbital case will not have a pairing instability upon projection onto the $j=3/2$ manifold due to mixing with the on-site spin singlet channel. We showed in Section~\ref{projectionconsequence} that under a similar projection onto a single band in the two orbital case, the VC state (Channel 1 of Table~\ref{table:1}) also becomes repulsive with interaction $U+U^\prime$ due to mixing with the on-site spin singlet channel (Channel 4). It becomes energetically stable only when both bands are allowed to cross the Fermi energy. By analogy, the VC analog state in the three orbital case becomes attractive when bands associated with both the $j=1/2$ and $j=3/2$ manifolds are allowed to cross the Fermi energy. 

\subsection{Other pairing channels and their stability}

In addition to the $A_{1g}$ pairing channel, we also find channels belonging to a $T_{2g}$ triplet (Channels B-D) and an $E_g$ doublet (Channels E and F). The pairing interactions of these channels can also be explained by mixing between spin triplet and spin singlet states of the zero spin-orbit coupling limit upon projection onto to the lower energy manifold. 

To illustrate this for the $T_{2g}$ channels, take $\hat{\Delta}_{0,T_{2g}}=\Delta_{0,T_{2g}}\lambda_1\otimes\sigma_0(i\sigma_2)$ as the spin singlet channel and $\hat{\Delta}_{1,T_{2g}}=\Delta_{1,T_{2g}}(\lambda_6\otimes\sigma_2+\lambda_5\otimes\sigma_1)(i\sigma_2)$ as the spin triplet channel. With respect to the atomic interactions, these two channels have interaction energies $g_{0,T_{2g}}=U^\prime+J$ and $g_{1,T_{2g}}=U^\prime-J$, respectively (Table~\ref{table:3}). Following the same procedure as for the $A_{1g}$ case, the linearized gap equations for $\Delta_{0,T_{2g}}$ and $\Delta_{1,T_{2g}}$ can be derived after projecting out the higher energy manifold:
\begin{align}
\frac{4\Delta_{0,T_{2g}}}{g_{0,T_{2g}}}&=-\sum_{\vec{k}}\frac{2}{3}(\Delta_{0,T_{2g}}-2\Delta_{1,T_{2g}})f(\epsilon_4) \nonumber \\
\frac{8\Delta_{1,T_{2g}}}{g_{1,T_{2g}}}&=\sum_{\vec{k}}\frac{4}{3}(\Delta_{0,T_{2g}}-2\Delta_{1,T_{2g}})f(\epsilon_4).
\label{eq:cubic6}
\end{align}
We obtain $\Delta_{1,T_{2g}}/g_{1,T_{2g}}=-\Delta_{0,T_{2g}}/g_{0,T_{2g}}$ again implying that there is only one distinct gap obeying the gap equation:
\begin{equation}
\Delta_{0,T_{2g}}=-\frac{1}{3}\sum_{\vec{k}}(g_{0,T_{2g}}+2g_{1,T_{2g}})\frac{\Delta_{0,T_{2g}}}{2\epsilon_4}\tanh(\beta \epsilon_4/2),
\label{eq:cubic7}
\end{equation}
governed by the pairing interaction $(g_{0,T_{2g}}+2g_{1,T_{2g}})/3=U^\prime-J/3$, in agreement with the energy found for the $T_{2g}$ channels.

For the $E_g$ channels, we can similarly take $\hat{\Delta}_{0,E_g}=\Delta_{0,E_g}\lambda_7\otimes\sigma_0(i\sigma_2)$ as the spin singlet channel and $\hat{\Delta}_{1,T_{2g}}=\Delta_{1,E_g}(\lambda_6\otimes\sigma_1-\lambda_5\otimes\sigma_2)(i\sigma_2)$ as the spin triplet channel. The two gaps then obey:
\begin{align}
\frac{4\Delta_{0,E_g}}{g_{0,E_g}}&=-\sum_{\vec{k}}\frac{2}{3}(\Delta_{0,E_g}+2\Delta_{1,E_g})f(\epsilon_4) \nonumber \\
\frac{8\Delta_{1,E_g}}{g_{1,E_g}}&=-\sum_{\vec{k}}\frac{4}{3}(\Delta_{0,E_g}+2\Delta_{1,E_g})f(\epsilon_4),
\label{eq:cubic8}
\end{align}
with $g_{0,E_g}=U-J$ and $g_{1,E_g}=U^\prime-J$. These equations reduce to:
\begin{equation}
\Delta_{0,E_g}=-\frac{1}{3}\sum_{\vec{k}}(g_{0,E_g}+2g_{1,E_g})\frac{\Delta_{0,E_g}}{2\epsilon_4}\tanh(\beta \epsilon_4/2).
\label{eq:cubic9}
\end{equation}
Thus, the $E_g$ channels are governed by the pairing interaction $(g_{0,E_g}+2g_{1,E_g})/3=U/3+2U^\prime/3-J$. 

We can speculate on the conditions under which these other orbital singlet, spin triplet channels belonging to non-trivial degenerate irreducible representations can develop a finite $T_c$. Again, by analogy with the work done in Ref.~\onlinecite{vafek1}, if no bands are projected out, then the suppression of the pairing channels due to mixing with repulsive spin singlet channels of the same symmetry may be avoided. The $T_{2g}$ channels (Channels B-D) are of particular interest because even after projecting out the higher energy manifold, the pairing interaction can never acquire contributions from the Hubbard $U$ energy (the spin singlet channels of $T_{2g}$ symmetry do not involve intraorbital pairing).
Finally, in principle there remains a set of spin triplet channels of $T_{1g}$ symmetry (the last set of states in Table~\ref{table:3}), but their pairing remains purely interband for the symmetry allowed normal state Hamiltonian and therefore cannot generate a finite $T_c$ in the weak coupling limit.

\section{Summary and discussions} \label{sec:summary}

In this article, we studied the possible local pairing channels that arise from atomic interactions in tetragonal lattices with $d_{xz}$ and $d_{yz}$ orbitals on each site, and in octahedral lattices with $d_{xz}$, $d_{yz}$, and $d_{xy}$ orbitals on each site. In the tetragonal, two orbital case we confirmed the existence of the VC state of $A_{1g}$ symmetry. The VC state is only stable when both bands are allowed to cross the chemical potential. In addition, a degenerate pair of states of $E_g$ symmetry also exist and they are stable regardless of how many bands cross the chemical potential. All three states have pairing interaction $U^\prime-J$ and can become attractive when $J>U^\prime.$ Such on-site Cooper pairs are necessarily even parity. 

Spin-orbit coupling plays a pivotal role in the realization of such novel ``spin triplet, orbital singlet'' superconducting states. Not only is its presence or absence important, but the form of spin-orbit coupling (here, whether it is local or non-local) also dictates whether the additional spin triplet pairing channels with non-trivial symmetry develop intraband pairing and can survive in the weak coupling regime. In particular, the stabilization of the channels with $E_g$ symmetry by including non-local spin-orbit interactions introduces another layer of novelty to the potential superconducting states. Superconducting channels belonging to degenerate, non-trivial $E_g$ symmetry have been proposed to exist in heavy fermion systems such URu$_2$Si$_2$ in which the pairs are even under parity~\cite{schemm1,yano1,kasahara1}.

It is interesting to note that in the strong coupling DMFT study of Ref.~\onlinecite{hoshino1}, anisotropy in the spin interactions was found to play a critical role in stabilizing the superconducting state. Another cause of spin anisotropy is of course single particle spin-orbit coupling. Thus, both strong and weak coupling studies point to spin-orbit coupling as a key partner to the Hund's interaction in this superconducting mechanism. It is worthwhile to extend the DMFT work of Ref.~\onlinecite{hoshino1} to include spin-orbit interactions in a more rigorous fashion. Moreover, non-local spin-orbit interactions must also be considered in light of our results. 

As in the two orbital tetragonal case, in the cubic, three orbital case, the analog of the VC state can only be stable when all bands are allowed to cross the chemical potential. This is also true for the $E_g$ and $T_{2g}$ pairing states in the cubic case, unlike in the tetragonal case in which the $E_g$ channel is generically stable (once $J>U^\prime$). When the $j=1/2$ manifold is projected out, the mixing between spin triplet and spin singlet states belonging to the same irreducible representations prevents the existence of channels with interaction $U^\prime-J$. Hence, in the search for materials which may realize such spin triplet, orbital singlet Cooper pairs such as the VC state, the number of bands crossing the Fermi energy must be considered. 

From a theoretical perspective, the full consequences of including orbital degrees of freedom in superconductivity are still not fully understood. Recent works continue to reveal new features such as emergent Bogoliubov Fermi surfaces~\cite{agterberg1} and topological non-trivial nodes in the gap function~\cite{brydon1,timm1}. From an experimental perspective, a natural next step would be to predict the experimental signatures of such novel spin triplet superconducting states in measurements of the spin susceptibility and Knight shift~\cite{abrikosov1,anderson1,frigeri1,yu1}. In addition to URu$_2$Si$_2$, Sr$_2$RuO$_4$~\cite{ishida1,mackenzie1} and SrTiO$_3$ for example are materials in which spin-orbit coupling is expected to be an important energy scale~\cite{haverkort1,swartz1}. It is also known that the bands associated with $d_{xz}$, $d_{yz}$, and $d_{xy}$ orbitals all cross the Fermi energy for these materials. It will be of interest to combine our results with more realistic microscopic descriptions of these materials.

\section*{Acknowledgments}

We thank S. Raghu, Y. Yu, P. Brydon, A. Ramires, and M. Sigrist for useful discussions. A. C. was supported by the DOE Office of Basic Energy Sciences, contract DEAC02- 76SF00515 and a NSERC PGS-D Scholarship. D. F. A. was supported by the Gordon and Betty Moore Foundation's EPiQS Initiative through Grant No. GBMF4302.

\appendix
\section{Orbital space operators in terms of Gell-Mann matrices} \label{appendix}

In this appendix, we clarify how our operators in orbital space are expressed in terms of the nine Gell-Mann matrices. The most general quadratic operator $\hat{\mathcal{O}}$ in an orbital space with three orbitals may be expressed as:
\begin{equation}
\hat{\mathcal{O}}=\sum_{\gamma\gamma^\prime=1}^3 c^\dag_{\gamma}\mathcal{O}_{\gamma\gamma^\prime}c_{\gamma^\prime},
\label{eq:ap1}
\end{equation}
where $\gamma/\gamma^\prime=1,2,3$ indexes the orbitals. Throughout this work, we have labeled the $d_{xz}$ orbital as $\gamma=1$, the $d_{yz}$ orbital as $\gamma=2$, and the $d_{xy}$ orbital as $\gamma=3$. If $\hat{\mathcal{O}}$ is Hermitian, then the $3\times 3$ $\mathcal{O}_{\gamma\gamma^\prime}$ matrix can be expressed as a linear combination of the nine Gell-Mann matrices $\lambda_\mu$:
\begin{equation}
\mathcal{O}_{\gamma\gamma^\prime}=\sum_{\mu=0}^8 A_\mu (\lambda_\mu)_{\gamma\gamma^\prime},
\label{eq:ap2}
\end{equation}
where $A_\mu$ are expansion coefficients. The Gell-Mann matrices we use are:
\begin{align}
\lambda_0&=\begin{pmatrix} 1&0&0\\ 0&1&0\\0&0&1 \end{pmatrix} \nonumber \\
\lambda_1&=\begin{pmatrix} 0&1&0\\ 1&0&0\\0&0&0 \end{pmatrix},\ \lambda_2=\begin{pmatrix} 0&0&1\\ 0&0&0\\1&0&0 \end{pmatrix},\ \lambda_3 = \begin{pmatrix} 0&0&0\\ 0&0&1\\0&1&0 \end{pmatrix} \nonumber \\
\lambda_4&=\begin{pmatrix} 0&-i&0\\ i&0&0\\0&0&0 \end{pmatrix},\ \lambda_5=\begin{pmatrix} 0&0&-i\\ 0&0&0\\i&0&0 \end{pmatrix},\ \lambda_6 = \begin{pmatrix} 0&0&0\\ 0&0&-i\\0&i&0 \end{pmatrix} \nonumber \\
\lambda_7&=\begin{pmatrix} 1&0&0\\ 0&-1&0\\0&0&0 \end{pmatrix},\ \lambda_8=\frac{1}{\sqrt{3}}\begin{pmatrix} 1&0&0\\ 0&1&0\\0&0&-2 \end{pmatrix}.
\label{eq:ap3}
\end{align}
The Gell-Mann matrices thus form a basis for the space of all Hermitian $3\times 3$ matrices just as the Pauli matrices form a basis for the space of all Hermitian $2\times 2$ matrices.

\bibliography{bibliography}{}

\begin{thebibliography}{28}
\expandafter\ifx\csname natexlab\endcsname\relax\def\natexlab#1{#1}\fi
\expandafter\ifx\csname bibnamefont\endcsname\relax
  \def\bibnamefont#1{#1}\fi
\expandafter\ifx\csname bibfnamefont\endcsname\relax
  \def\bibfnamefont#1{#1}\fi
\expandafter\ifx\csname citenamefont\endcsname\relax
  \def\citenamefont#1{#1}\fi
\expandafter\ifx\csname url\endcsname\relax
  \def\url#1{\texttt{#1}}\fi
\expandafter\ifx\csname urlprefix\endcsname\relax\def\urlprefix{URL }\fi
\providecommand{\bibinfo}[2]{#2}
\providecommand{\eprint}[2][]{\url{#2}}

\bibitem[{\citenamefont{Raghu et~al.}(2010)\citenamefont{Raghu, Kivelson, and
  Scalapino}}]{raghu1}
\bibinfo{author}{\bibfnamefont{S.}~\bibnamefont{Raghu}},
  \bibinfo{author}{\bibfnamefont{S.~A.} \bibnamefont{Kivelson}},
  \bibnamefont{and} \bibinfo{author}{\bibfnamefont{D.~J.}
  \bibnamefont{Scalapino}}, \bibinfo{journal}{Phys. Rev. B}
  \textbf{\bibinfo{volume}{81}}, \bibinfo{pages}{224505}
  (\bibinfo{year}{2010}),
  \urlprefix\url{https://link.aps.org/doi/10.1103/PhysRevB.81.224505}.

\bibitem[{\citenamefont{Kohn and Luttinger}(1965)}]{kohn1}
\bibinfo{author}{\bibfnamefont{W.}~\bibnamefont{Kohn}} \bibnamefont{and}
  \bibinfo{author}{\bibfnamefont{J.~M.} \bibnamefont{Luttinger}},
  \bibinfo{journal}{Phys. Rev. Lett.} \textbf{\bibinfo{volume}{15}},
  \bibinfo{pages}{524} (\bibinfo{year}{1965}),
  \urlprefix\url{https://link.aps.org/doi/10.1103/PhysRevLett.15.524}.

\bibitem[{\citenamefont{Dagotto et~al.}(2001)\citenamefont{Dagotto, Hotta, and
  Moreo}}]{dagotto1}
\bibinfo{author}{\bibfnamefont{E.}~\bibnamefont{Dagotto}},
  \bibinfo{author}{\bibfnamefont{T.}~\bibnamefont{Hotta}}, \bibnamefont{and}
  \bibinfo{author}{\bibfnamefont{A.}~\bibnamefont{Moreo}},
  \bibinfo{journal}{Physics Reports} \textbf{\bibinfo{volume}{344}},
  \bibinfo{pages}{1} (\bibinfo{year}{2001}).

\bibitem[{\citenamefont{Spa\l{}ek}(2001)}]{spalek1}
\bibinfo{author}{\bibfnamefont{J.}~\bibnamefont{Spa\l{}ek}},
  \bibinfo{journal}{Phys. Rev. B} \textbf{\bibinfo{volume}{63}},
  \bibinfo{pages}{104513} (\bibinfo{year}{2001}),
  \urlprefix\url{https://link.aps.org/doi/10.1103/PhysRevB.63.104513}.

\bibitem[{\citenamefont{Han}(2004)}]{han1}
\bibinfo{author}{\bibfnamefont{J.~E.} \bibnamefont{Han}},
  \bibinfo{journal}{Phys. Rev. B} \textbf{\bibinfo{volume}{70}},
  \bibinfo{pages}{054513} (\bibinfo{year}{2004}),
  \urlprefix\url{https://link.aps.org/doi/10.1103/PhysRevB.70.054513}.

\bibitem[{\citenamefont{Klejnberg and Spalek}(1999)}]{klejnberg1}
\bibinfo{author}{\bibfnamefont{A.}~\bibnamefont{Klejnberg}} \bibnamefont{and}
  \bibinfo{author}{\bibfnamefont{J.}~\bibnamefont{Spalek}},
  \bibinfo{journal}{Journal of Physics: Condensed Matter}
  \textbf{\bibinfo{volume}{11}}, \bibinfo{pages}{6553} (\bibinfo{year}{1999}).

\bibitem[{\citenamefont{Hoshino and Werner}(2015)}]{hoshino1}
\bibinfo{author}{\bibfnamefont{S.}~\bibnamefont{Hoshino}} \bibnamefont{and}
  \bibinfo{author}{\bibfnamefont{P.}~\bibnamefont{Werner}},
  \bibinfo{journal}{Phys. Rev. Lett.} \textbf{\bibinfo{volume}{115}},
  \bibinfo{pages}{247001} (\bibinfo{year}{2015}),
  \urlprefix\url{https://link.aps.org/doi/10.1103/PhysRevLett.115.247001}.

\bibitem[{\citenamefont{Werner et~al.}(2008)\citenamefont{Werner, Gull, Troyer,
  and Millis}}]{werner1}
\bibinfo{author}{\bibfnamefont{P.}~\bibnamefont{Werner}},
  \bibinfo{author}{\bibfnamefont{E.}~\bibnamefont{Gull}},
  \bibinfo{author}{\bibfnamefont{M.}~\bibnamefont{Troyer}}, \bibnamefont{and}
  \bibinfo{author}{\bibfnamefont{A.~J.} \bibnamefont{Millis}},
  \bibinfo{journal}{Phys. Rev. Lett.} \textbf{\bibinfo{volume}{101}},
  \bibinfo{pages}{166405} (\bibinfo{year}{2008}),
  \urlprefix\url{https://link.aps.org/doi/10.1103/PhysRevLett.101.166405}.

\bibitem[{\citenamefont{Vafek and Chubukov}(2017)}]{vafek1}
\bibinfo{author}{\bibfnamefont{O.}~\bibnamefont{Vafek}} \bibnamefont{and}
  \bibinfo{author}{\bibfnamefont{A.~V.} \bibnamefont{Chubukov}},
  \bibinfo{journal}{Phys. Rev. Lett.} \textbf{\bibinfo{volume}{118}},
  \bibinfo{pages}{087003} (\bibinfo{year}{2017}),
  \urlprefix\url{https://link.aps.org/doi/10.1103/PhysRevLett.118.087003}.

\bibitem[{\citenamefont{Fernandes and Vafek}(2014)}]{fernandes1}
\bibinfo{author}{\bibfnamefont{R.~M.} \bibnamefont{Fernandes}}
  \bibnamefont{and} \bibinfo{author}{\bibfnamefont{O.}~\bibnamefont{Vafek}},
  \bibinfo{journal}{Phys. Rev. B} \textbf{\bibinfo{volume}{90}},
  \bibinfo{pages}{214514} (\bibinfo{year}{2014}),
  \urlprefix\url{https://link.aps.org/doi/10.1103/PhysRevB.90.214514}.

\bibitem[{\citenamefont{Ramires et~al.}(2018)\citenamefont{Ramires, Agterberg,
  and Sigrist}}]{ramires1}
\bibinfo{author}{\bibfnamefont{A.}~\bibnamefont{Ramires}},
  \bibinfo{author}{\bibfnamefont{D.~F.} \bibnamefont{Agterberg}},
  \bibnamefont{and} \bibinfo{author}{\bibfnamefont{M.}~\bibnamefont{Sigrist}},
  \bibinfo{journal}{Phys. Rev. B} \textbf{\bibinfo{volume}{98}},
  \bibinfo{pages}{024501} (\bibinfo{year}{2018}),
  \urlprefix\url{https://link.aps.org/doi/10.1103/PhysRevB.98.024501}.

\bibitem[{\citenamefont{Ramires and Sigrist}(2016)}]{ramires2}
\bibinfo{author}{\bibfnamefont{A.}~\bibnamefont{Ramires}} \bibnamefont{and}
  \bibinfo{author}{\bibfnamefont{M.}~\bibnamefont{Sigrist}},
  \bibinfo{journal}{Phys. Rev. B} \textbf{\bibinfo{volume}{94}},
  \bibinfo{pages}{104501} (\bibinfo{year}{2016}),
  \urlprefix\url{https://link.aps.org/doi/10.1103/PhysRevB.94.104501}.

\bibitem[{\citenamefont{Schemm et~al.}(2015)\citenamefont{Schemm, Baumbach,
  Tobash, Ronning, Bauer, and Kapitulnik}}]{schemm1}
\bibinfo{author}{\bibfnamefont{E.~R.} \bibnamefont{Schemm}},
  \bibinfo{author}{\bibfnamefont{R.~E.} \bibnamefont{Baumbach}},
  \bibinfo{author}{\bibfnamefont{P.~H.} \bibnamefont{Tobash}},
  \bibinfo{author}{\bibfnamefont{F.}~\bibnamefont{Ronning}},
  \bibinfo{author}{\bibfnamefont{E.~D.} \bibnamefont{Bauer}}, \bibnamefont{and}
  \bibinfo{author}{\bibfnamefont{A.}~\bibnamefont{Kapitulnik}},
  \bibinfo{journal}{Phys. Rev. B} \textbf{\bibinfo{volume}{91}},
  \bibinfo{pages}{140506} (\bibinfo{year}{2015}),
  \urlprefix\url{https://link.aps.org/doi/10.1103/PhysRevB.91.140506}.

\bibitem[{\citenamefont{Yano et~al.}(2008)\citenamefont{Yano, Sakakibara,
  Tayama, Yokoyama, Amitsuka, Homma, Miranovi\ifmmode~\acute{c}\else
  \'{c}\fi{}, Ichioka, Tsutsumi, and Machida}}]{yano1}
\bibinfo{author}{\bibfnamefont{K.}~\bibnamefont{Yano}},
  \bibinfo{author}{\bibfnamefont{T.}~\bibnamefont{Sakakibara}},
  \bibinfo{author}{\bibfnamefont{T.}~\bibnamefont{Tayama}},
  \bibinfo{author}{\bibfnamefont{M.}~\bibnamefont{Yokoyama}},
  \bibinfo{author}{\bibfnamefont{H.}~\bibnamefont{Amitsuka}},
  \bibinfo{author}{\bibfnamefont{Y.}~\bibnamefont{Homma}},
  \bibinfo{author}{\bibfnamefont{P.}~\bibnamefont{Miranovi\ifmmode~\acute{c}\else
  \'{c}\fi{}}}, \bibinfo{author}{\bibfnamefont{M.}~\bibnamefont{Ichioka}},
  \bibinfo{author}{\bibfnamefont{Y.}~\bibnamefont{Tsutsumi}}, \bibnamefont{and}
  \bibinfo{author}{\bibfnamefont{K.}~\bibnamefont{Machida}},
  \bibinfo{journal}{Phys. Rev. Lett.} \textbf{\bibinfo{volume}{100}},
  \bibinfo{pages}{017004} (\bibinfo{year}{2008}),
  \urlprefix\url{https://link.aps.org/doi/10.1103/PhysRevLett.100.017004}.

\bibitem[{\citenamefont{Kasahara et~al.}(2007)\citenamefont{Kasahara, Iwasawa,
  Shishido, Shibauchi, Behnia, Haga, Matsuda, Onuki, Sigrist, and
  Matsuda}}]{kasahara1}
\bibinfo{author}{\bibfnamefont{Y.}~\bibnamefont{Kasahara}},
  \bibinfo{author}{\bibfnamefont{T.}~\bibnamefont{Iwasawa}},
  \bibinfo{author}{\bibfnamefont{H.}~\bibnamefont{Shishido}},
  \bibinfo{author}{\bibfnamefont{T.}~\bibnamefont{Shibauchi}},
  \bibinfo{author}{\bibfnamefont{K.}~\bibnamefont{Behnia}},
  \bibinfo{author}{\bibfnamefont{Y.}~\bibnamefont{Haga}},
  \bibinfo{author}{\bibfnamefont{T.~D.} \bibnamefont{Matsuda}},
  \bibinfo{author}{\bibfnamefont{Y.}~\bibnamefont{Onuki}},
  \bibinfo{author}{\bibfnamefont{M.}~\bibnamefont{Sigrist}}, \bibnamefont{and}
  \bibinfo{author}{\bibfnamefont{Y.}~\bibnamefont{Matsuda}},
  \bibinfo{journal}{Phys. Rev. Lett.} \textbf{\bibinfo{volume}{99}},
  \bibinfo{pages}{116402} (\bibinfo{year}{2007}),
  \urlprefix\url{https://link.aps.org/doi/10.1103/PhysRevLett.99.116402}.

\bibitem[{\citenamefont{Ikeda et~al.}(2012)\citenamefont{Ikeda, Suzuki, Arita,
  Takimoto, Shibauchi, and Matsuda}}]{ikeda1}
\bibinfo{author}{\bibfnamefont{H.}~\bibnamefont{Ikeda}},
  \bibinfo{author}{\bibfnamefont{M.-T.} \bibnamefont{Suzuki}},
  \bibinfo{author}{\bibfnamefont{R.}~\bibnamefont{Arita}},
  \bibinfo{author}{\bibfnamefont{T.}~\bibnamefont{Takimoto}},
  \bibinfo{author}{\bibfnamefont{T.}~\bibnamefont{Shibauchi}},
  \bibnamefont{and} \bibinfo{author}{\bibfnamefont{Y.}~\bibnamefont{Matsuda}},
  \bibinfo{journal}{Nature Physics} \textbf{\bibinfo{volume}{8}},
  \bibinfo{pages}{528} (\bibinfo{year}{2012}).

\bibitem[{\citenamefont{Agterberg et~al.}(2017)\citenamefont{Agterberg, Brydon,
  and Timm}}]{agterberg1}
\bibinfo{author}{\bibfnamefont{D.~F.} \bibnamefont{Agterberg}},
  \bibinfo{author}{\bibfnamefont{P.~M.~R.} \bibnamefont{Brydon}},
  \bibnamefont{and} \bibinfo{author}{\bibfnamefont{C.}~\bibnamefont{Timm}},
  \bibinfo{journal}{Phys. Rev. Lett.} \textbf{\bibinfo{volume}{118}},
  \bibinfo{pages}{127001} (\bibinfo{year}{2017}),
  \urlprefix\url{https://link.aps.org/doi/10.1103/PhysRevLett.118.127001}.

\bibitem[{\citenamefont{Ho and Yip}(1999)}]{ho1}
\bibinfo{author}{\bibfnamefont{T.-L.} \bibnamefont{Ho}} \bibnamefont{and}
  \bibinfo{author}{\bibfnamefont{S.}~\bibnamefont{Yip}},
  \bibinfo{journal}{Phys. Rev. Lett.} \textbf{\bibinfo{volume}{82}},
  \bibinfo{pages}{247} (\bibinfo{year}{1999}),
  \urlprefix\url{https://link.aps.org/doi/10.1103/PhysRevLett.82.247}.

\bibitem[{\citenamefont{Brydon et~al.}(2016)\citenamefont{Brydon, Wang,
  Weinert, and Agterberg}}]{brydon1}
\bibinfo{author}{\bibfnamefont{P.~M.~R.} \bibnamefont{Brydon}},
  \bibinfo{author}{\bibfnamefont{L.}~\bibnamefont{Wang}},
  \bibinfo{author}{\bibfnamefont{M.}~\bibnamefont{Weinert}}, \bibnamefont{and}
  \bibinfo{author}{\bibfnamefont{D.~F.} \bibnamefont{Agterberg}},
  \bibinfo{journal}{Phys. Rev. Lett.} \textbf{\bibinfo{volume}{116}},
  \bibinfo{pages}{177001} (\bibinfo{year}{2016}),
  \urlprefix\url{https://link.aps.org/doi/10.1103/PhysRevLett.116.177001}.

\bibitem[{\citenamefont{Timm et~al.}(2017)\citenamefont{Timm, Schnyder,
  Agterberg, and Brydon}}]{timm1}
\bibinfo{author}{\bibfnamefont{C.}~\bibnamefont{Timm}},
  \bibinfo{author}{\bibfnamefont{A.~P.} \bibnamefont{Schnyder}},
  \bibinfo{author}{\bibfnamefont{D.~F.} \bibnamefont{Agterberg}},
  \bibnamefont{and} \bibinfo{author}{\bibfnamefont{P.~M.~R.}
  \bibnamefont{Brydon}}, \bibinfo{journal}{Phys. Rev. B}
  \textbf{\bibinfo{volume}{96}}, \bibinfo{pages}{094526}
  (\bibinfo{year}{2017}),
  \urlprefix\url{https://link.aps.org/doi/10.1103/PhysRevB.96.094526}.

\bibitem[{\citenamefont{Abrikosov and Gor’kov}(1962)}]{abrikosov1}
\bibinfo{author}{\bibfnamefont{A.}~\bibnamefont{Abrikosov}} \bibnamefont{and}
  \bibinfo{author}{\bibfnamefont{L.}~\bibnamefont{Gor’kov}},
  \bibinfo{journal}{Sov. Phys. JETP} \textbf{\bibinfo{volume}{15}},
  \bibinfo{pages}{752} (\bibinfo{year}{1962}).

\bibitem[{\citenamefont{Anderson}(1959)}]{anderson1}
\bibinfo{author}{\bibfnamefont{P.~W.} \bibnamefont{Anderson}},
  \bibinfo{journal}{Phys. Rev. Lett.} \textbf{\bibinfo{volume}{3}},
  \bibinfo{pages}{325} (\bibinfo{year}{1959}),
  \urlprefix\url{https://link.aps.org/doi/10.1103/PhysRevLett.3.325}.

\bibitem[{\citenamefont{Frigeri et~al.}(2004)\citenamefont{Frigeri, Agterberg,
  Koga, and Sigrist}}]{frigeri1}
\bibinfo{author}{\bibfnamefont{P.~A.} \bibnamefont{Frigeri}},
  \bibinfo{author}{\bibfnamefont{D.~F.} \bibnamefont{Agterberg}},
  \bibinfo{author}{\bibfnamefont{A.}~\bibnamefont{Koga}}, \bibnamefont{and}
  \bibinfo{author}{\bibfnamefont{M.}~\bibnamefont{Sigrist}},
  \bibinfo{journal}{Phys. Rev. Lett.} \textbf{\bibinfo{volume}{92}},
  \bibinfo{pages}{097001} (\bibinfo{year}{2004}),
  \urlprefix\url{https://link.aps.org/doi/10.1103/PhysRevLett.92.097001}.

\bibitem[{\citenamefont{Yu et~al.}(2018)\citenamefont{Yu, Cheung, Raghu, and
  Agterberg}}]{yu1}
\bibinfo{author}{\bibfnamefont{Y.}~\bibnamefont{Yu}},
  \bibinfo{author}{\bibfnamefont{A.~K.} \bibnamefont{Cheung}},
  \bibinfo{author}{\bibfnamefont{S.}~\bibnamefont{Raghu}}, \bibnamefont{and}
  \bibinfo{author}{\bibfnamefont{D.}~\bibnamefont{Agterberg}},
  \bibinfo{journal}{arXiv preprint arXiv:1808.08029}  (\bibinfo{year}{2018}).

\bibitem[{\citenamefont{Ishida et~al.}(1998)\citenamefont{Ishida, Mukuda,
  Kitaoka, Asayama, Mao, Mori, and Maeno}}]{ishida1}
\bibinfo{author}{\bibfnamefont{K.}~\bibnamefont{Ishida}},
  \bibinfo{author}{\bibfnamefont{H.}~\bibnamefont{Mukuda}},
  \bibinfo{author}{\bibfnamefont{Y.}~\bibnamefont{Kitaoka}},
  \bibinfo{author}{\bibfnamefont{K.}~\bibnamefont{Asayama}},
  \bibinfo{author}{\bibfnamefont{Z.}~\bibnamefont{Mao}},
  \bibinfo{author}{\bibfnamefont{Y.}~\bibnamefont{Mori}}, \bibnamefont{and}
  \bibinfo{author}{\bibfnamefont{Y.}~\bibnamefont{Maeno}},
  \bibinfo{journal}{Nature} \textbf{\bibinfo{volume}{396}},
  \bibinfo{pages}{658} (\bibinfo{year}{1998}).

\bibitem[{\citenamefont{Mackenzie and Maeno}(2003)}]{mackenzie1}
\bibinfo{author}{\bibfnamefont{A.~P.} \bibnamefont{Mackenzie}}
  \bibnamefont{and} \bibinfo{author}{\bibfnamefont{Y.}~\bibnamefont{Maeno}},
  \bibinfo{journal}{Reviews of Modern Physics} \textbf{\bibinfo{volume}{75}},
  \bibinfo{pages}{657} (\bibinfo{year}{2003}).

\bibitem[{\citenamefont{Haverkort et~al.}(2008)\citenamefont{Haverkort,
  Elfimov, Tjeng, Sawatzky, and Damascelli}}]{haverkort1}
\bibinfo{author}{\bibfnamefont{M.~W.} \bibnamefont{Haverkort}},
  \bibinfo{author}{\bibfnamefont{I.~S.} \bibnamefont{Elfimov}},
  \bibinfo{author}{\bibfnamefont{L.~H.} \bibnamefont{Tjeng}},
  \bibinfo{author}{\bibfnamefont{G.~A.} \bibnamefont{Sawatzky}},
  \bibnamefont{and}
  \bibinfo{author}{\bibfnamefont{A.}~\bibnamefont{Damascelli}},
  \bibinfo{journal}{Phys. Rev. Lett.} \textbf{\bibinfo{volume}{101}},
  \bibinfo{pages}{026406} (\bibinfo{year}{2008}),
  \urlprefix\url{https://link.aps.org/doi/10.1103/PhysRevLett.101.026406}.

\bibitem[{\citenamefont{Swartz et~al.}(2018)\citenamefont{Swartz, Cheung, Yoon,
  Chen, Hikita, Raghu, and Hwang}}]{swartz1}
\bibinfo{author}{\bibfnamefont{A.~G.} \bibnamefont{Swartz}},
  \bibinfo{author}{\bibfnamefont{A.~K.~C.} \bibnamefont{Cheung}},
  \bibinfo{author}{\bibfnamefont{H.}~\bibnamefont{Yoon}},
  \bibinfo{author}{\bibfnamefont{Z.}~\bibnamefont{Chen}},
  \bibinfo{author}{\bibfnamefont{Y.}~\bibnamefont{Hikita}},
  \bibinfo{author}{\bibfnamefont{S.}~\bibnamefont{Raghu}}, \bibnamefont{and}
  \bibinfo{author}{\bibfnamefont{H.~Y.} \bibnamefont{Hwang}},
  \bibinfo{journal}{arXiv preprint arXiv:1805.00047}  (\bibinfo{year}{2018}).

\end{thebibliography}

\end{document}